\documentclass[12pt]{iopart}
\usepackage{graphicx}

\newcommand{\Red}[1]{#1}

\renewcommand{\eta}{\tau}
\newcommand{\ds}{\displaystyle}

\begin{document}
\title{Buoyancy and Penrose Process Produce Jets from Rotating Black Holes}
\author{V S Semenov$^1$, S A Dyadechkin$^{1,2}$, M F Heyn$^3$}
\address{$^1$ St.~Petersburg State University, Petrodvoretz, 198504, Russia}
\address{$^2$ Finish Meteorological Institute, Helsinki, Finland}
\address{$^3$ Institute for Theoretical and Computational Physics, TU-Graz, Austria}

\begin{abstract}
{}
{The exact mechanism by which astrophysical jets are formed is still unknown.
It is believed that necessary elements are a rotating (Kerr) black hole and a
magnetised accreting plasma.}
{We model the accreting plasma as a
collection of magnetic flux tubes/strings. If such a tube falls into a Kerr
black hole, then the leading portion loses angular momentum and energy
as the string brakes, and to compensate for this loss, momentum and
energy is redistributed to the trailing portion of the tube.}
{We found that buoyancy creates a pronounced helical magnetic field
structure aligned with the spin axis. Along the field lines, the plasma is
centrifugally accelerated close to the speed of light. This process
leads to unlimited stretching of the flux tube since one part of the tube
continues to fall into the black hole and simultaneously the other part
of the string is pushed outward.
Eventually, reconnection cuts the tube, the inner part is filled with
new material and the outer part forms a collimated bubble-structured
relativistic jet.
Each plasmoid can be considered like an outgoing particle in the Penrose
mechanism: it carries extracted rotational energy away from the black hole
while the falling part with the corresponding negative energy
is left inside the ergosphere.}
\bigskip

\hspace*{-2cm}\normalsize Published in: {\it Physica scripta {\bf 89}, 045003, 2014.}
\end{abstract}

\maketitle


%
\section{Introduction}
Astrophysical jet streams can be found all over the universe and
are manifestations of violent energetic outbursts from massive
cosmic objects. Those jets are a mixture of super heated, low
density gas, extremely energetic particles and magnetic fields
which are ejected from the pole area in the form of narrow columns
of gas saturated with elementary particles. On the intergalactic
scale, these powerful emissions originate in the core of active
galactic nuclei (AGNs), including exotic astrophysical objects
such as quasars and
blazars~\cite{Rawlings1991-138,Narayan2005-199,Giannios2009-L29}.

In recent years analytical
work~\cite{Blandford1977-433,Narayan2005-199,Punsly2001,Beskin2009}
and numerical
simulations~\cite{Koide2002-1688,McKinney2006-1561,Hawley2007-117,Hawley2006-103}
reveal  ingredients which are necessary to form a jet: plasma
accretion, magnetic field and a black hole. To underline the physics
it is interesting to look at the interaction of the magnetic
field with the accreting matter drawn towards the black hole and
its role in the jet formation process. This interaction will be
most effective in the presence of a rotating (Kerr) black
hole~\cite{Misner1973}. Two different models of energy extraction
from a rotating black hole are often discussed in the literature.
On one hand the Penrose mechanism where inside the ergosphere a
particle splits into two particles, one possibly escapes to infinity
with higher mass-energy than the original infalling particle,
whilst the other particle falls with negative
mass-energy past the event horizon into the
hole~\cite{Penrose1969-252,Penrose1971-177}. This mechanism is
based on frame-dragging of rotating bodies as first derived within
general relativity by Lense and Thirring in
1918~\cite{Lense1918-156}. On the other hand the Blandford-Znajek
mechanism~\cite{Blandford1977-433} where the accreted material has
a magnetic field threaded through as it falls into the rotating
black hole. Again frame-dragging coils and twists the magnetic
field like a rope. The associated huge electric currents will
eventually transfer their energy into the plasma which is blown
away from the black hole in form of
jets~\cite{Punsly2001,Beskin2009}.

\Red{ Because of frame-dragging there exists a region around the
Kerr black hole called the ergosphere, the outer surface is the
static limit boundary.
It is known that the energy of a particle can assume both
positive and negative values inside the ergosphere as measured by
an observer at infinity~\cite{Misner1973}. 
If a particle with zero
angular momentum falls into the black hole, an external observer
will see the particle to start to  corotate together with the
black hole. The deeper the particle falls into the black hole, the
faster it will be seen to rotate from the external observer.
Positive and negative angular momentum for a particular particle
can be defined if it rotates faster or slower than the zero
angular particle as it is seen from an external observer. The
rotation induced by frame-dragging tends to zero at infinity and
takes a maximal value at the event horizon, so it constitutes a
strong differential rotation. }

A magnetised plasma can be modelled as a gas-like collection of
magnetic flux tubes, each of them behaves as a nonlinear string.
Therefore, one can simulate the motion of a flux tube (2D problem)
instead of solving numerically the complete set of MHD equations
(4D problem). This approach has been successfully applied to the
model solar cycle (\cite{babcock1961-572}), the magnetic barrier
(or depletion layer) at the dayside magnetopause
(\cite{zwan1976-1636,erkaev2006-209}), and magnetic reconnection
in the magnetotail (\cite{pudovkin1985-1, chen1999-14613}). This
method has also been used in the problem of relativistic string -
Kerr black hole interaction: in (Semenov, 2000) the general
problem was stated, subsequent first numerical simulation showed
the possibility of energy extraction with the help of a flux tube
from a rotating black hole (Semenov, 2002), and finally an
improved numerical scheme was able to simulate the appearance of
the jet (Semenov 2004). In the present paper the important role of
buoyancy and magnetic reconnection is discussed in the process of
jet formation as well as the propagation of plasmoids outwards the
black hole.

\renewcommand{\eta}{\tau}

\section{\Red{Thin flux tube approximation}}
\begin{figure}
\centering
\includegraphics[width=0.3\hsize]{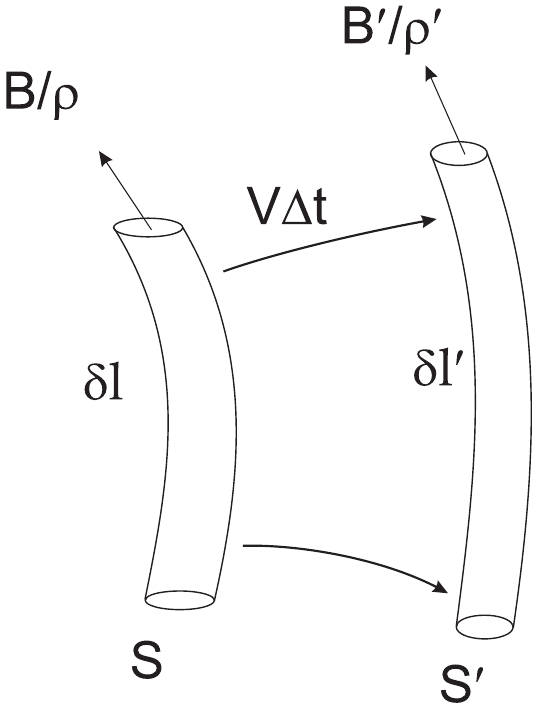}
\caption{\Red{The frozen-in motion of a flux tube together with the plasma.}
}
\label{fig_0}
\end{figure}

\Red{ 
In classical ideal Magnetohydrodynamics (MHD), the magnetic
field is frozen into the flow, i.e.\ magnetic flux tubes move
together with the flow as sketched in Fig.~\ref{fig_0}; 
if a flux tube connects two elements of
plasma it will do so during the evolution of the plasma
flow~\cite{Landau1984}. 
From a mathematical point of view
this means that the Lie derivative of the two vector fields, 
namely the ratio of magnetic field over plasma density $\mathbf{B}/\rho$ 
and plasma velocity $\mathbf{v}$,
vanishes~\cite{pudovkin1985-1} and one can use this set of vector fields as
basis vectors for a coordinate system where the
fluid trajectories (parametrised by $\tau$) and the magnetic field
lines (parametrised by $\alpha$) serve as coordinate lines~\cite{Misner1973}. 
In those coordinates the convective derivative and the the magnetic
stress term in the momentum equation become simple,
\begin{equation}
\frac{\partial^2 \bf r }{\partial \tau^2}-\frac{1}{4\pi
}\frac{\partial}{\partial \alpha}\left(\frac{\partial({\rho \bf r
})}{\partial \alpha} \right)=
-\frac{1}{\rho}\nabla{P}(\mathbf{r})-\nabla{\Phi}.
\label{motiona_s}
\end{equation}
Here, ${\bf r}(\tau,\alpha)$ is the position of a point on the
flux tube, $\rho$ is the plasma density, $P$ is the total pressure
(gas plus magnetic), and $\Phi$ is the gravitational potential.
Once the function ${\bf r}(\tau,\alpha)$ is found, 
the velocity field  and magnetic field are obtained from 
\begin{equation}
\frac{\partial{\mathbf{r}}}{\partial{\tau}} ={\mathbf{v}},
\hspace{12pt}
\frac{\partial{\mathbf{r}}}{\partial{\alpha}}=\frac{\mathbf{B}}{\rho}.
 \label{rel2}
\end{equation}
If one considers the right hand side of the equation
(\ref{motiona_s}), i.e. the distribution of the total pressure and
the gravitational potential as a given functions, this equation is
similar to the equation of the nonlinear elastic string and it
describes the dynamics of a massive infinitely thin flux tube
embedded in a plasma. Note, that the plasma density can be found
from the following nonlinear algebraic equation which is just
definition of the total pressure
\begin{equation}
P(\mathbf{r})=p(\rho)+\frac{\rho^2}{8\pi}\left(\frac{\partial{ \bf
r }}{\partial \alpha} \right)^2,
 \label{totalP}
\end{equation}
where $p(\rho)$ is the equation of state, for example an adiabatic
law.

Equation set (\ref{motiona_s}) governs plasma
inertia, magnetic tension forces, the redistribution of energy,
and buoyancy effects. It is a hyperbolic system which describes
slow and Alf\'enic MHD waves in a given total pressure gradient
distribution, i.e.\ the accumulation and relaxation of Maxwellian
stresses. By specifying the total pressure, the fast waves
dynamics is not selfconsistently described any more. But the
essential physics is still kept as the remaining waves are able to
redistribute angular momentum and energy along the massive flux
tube which by itself conserves the line integrated total angular
momentum and energy.

The buoyancy of a thin flux tube can be taken into account
like it has been shown in Ref.~\cite{Parker1979} .
For this, let us consider a flux tube embedded in a
plasma without magnetic field under static equilibrium
\begin{equation}
\frac{1}{\rho^*}\nabla{p(\rho^*)}(\mathbf{r})+\nabla{\Phi}=0.
\label{equilibrium}
\end{equation}
The distribution of the gas pressure outside the tube can be used
as the total pressure distribution in Equation
(\ref{motiona_s}),
$p(\rho^*)=p(\mathbf{r})=P(\mathbf{r})$,
and this equation takes the form
\begin{equation}
\frac{\partial^2 \bf r }{\partial \tau^2}-\frac{1}{4\pi
}\frac{\partial}{\partial \alpha}\left(\frac{\partial({\rho \bf r
})}{\partial \alpha} \right)=
-(\frac{1}{\rho}-\frac{1}{\rho^*})\nabla{P}(\mathbf{r}),
\label{buoyancy}
\end{equation}
where $\rho^*$ is the plasma density outside the flux tube.
Therefore if $\rho<\rho^*$ the tube will rise whereas for
$\rho>\rho^*$ the flux tube will sink.
The magnetic pressure inside the tube compensates part of the
total pressure in Equation (\ref{equilibrium}).
This means that the magnetic field stimulates the
buoyancy of the tube: the stronger the magnetic field, the
faster the tube emerges.

It is interesting to remember that buoyancy of magnetic tubes is
believed to be responsible for the formation of the solar
spots~\cite{babcock1961-572}. Hence, the buoyancy effect is a
necessary element for solar cycle theory.

\section{Thin flux tubes in the relativistic MHD}

 The relativistic magnetohydrodynamic (RMHD) equations can be
presented in terms of the time-like vector of the 4-velocity
$u^i$, $u^i u_i=1$ and the space-like 4-vector of the magnetic
field
\begin{equation}
 h^i=\ast F^{ik}u_k, \label{h}
\end{equation}
where $\ast F^{ik}$ is the dual tensor of the electromagnetic
field,  $h^i h_i<0$, ~\cite{Lichnerowicz}
\begin{eqnarray}
&\ds\nabla_i\rho u^i  =  0, \label{1}
\\
&\ds\nabla_i T^{ik}  =  0, \label{2}
 \\
&\ds\nabla_i(h^i u^k-h^k u^i)  =  0. \label{3}
\end{eqnarray}
Here,  $T^{ij}$ is the stress-energy tensor,
\begin{eqnarray}
&T^{ij}=Q\, u^i u^j - P\, g^{ij}-\frac {1}{4\pi}\, h^i h^j, \label{4}
\end{eqnarray}
where
\begin{eqnarray}
P \equiv p-\frac {1}{8\pi}\,h^k h_k,\quad Q \equiv
p+\varepsilon-\frac{1}{4\pi}\,h^k h_k, \label{11}
\end{eqnarray}
 $p$ is the plasma pressure, $P$ is the total (plasma plus
magnetic) pressure, $\varepsilon $ is the internal energy
including $\rho c^2$, and $g_{ik}$ is the metric tensor with
signature $(1,-1,-1,-1)$.

Equation (\ref{1}) is the continuity equation, (\ref{2}) are the
energy-momentum equations, and (\ref{3}) are  Maxwell equations.

Using (\ref{1}) Maxwell equations (\ref{3})  can be rewritten in
the form of a Lie derivative
\begin{equation}
\frac{h^i}{\rho}\nabla_i \frac{u^k}{q}= \frac{u^i}{q}\nabla_i
\frac{h^k}{\rho},\label{6}
\end{equation}
and one can introduce coordinates $\eta, \alpha$ such that~\cite{Misner1973}
\begin{eqnarray}
x^i_\eta \equiv \frac{\partial {x^i}}{\partial \eta} =
\frac{u^i}{q}, \quad
\ x^i_\alpha \equiv \frac{\partial{x^i}}{\partial \alpha} = \frac{h^i}{\rho},
 \label{7}
\end{eqnarray}
with new coordinate vectors $u^i/q$, $h^i/\rho$ tangent to the
trajectory of a fluid element and to the magnetic field line. This
allows to define a magnetic flux tube as a bundle of the magnetic
field lines $h^i/\rho$. The motion of the flux tube is introduced
as a Lie dragging of the vector field $h^i/\rho$ along the vector
field $u^i/q$.
It turns out \cite{Semenov2002-13} that the mass coordinate
$\alpha$ along the  flux tube has the sense of the mass of the
plasma for a tube with unit flux in the proper system of
reference. The second coordinate $\eta$ is not  Lagrangian or
proper time, but is just a time-like parameter which traces the
flux tube in space-time of general relativity.

Using (\ref{7}), the energy-momentum equation (\ref{2}) can be
rearranged to form a set of string equations
\begin{equation}
 -\frac{\partial }{\partial \eta}
 \left(\frac{Q q}{\rho}x^l_\eta\right)-
 \frac{Q q}{\rho}\Gamma^l_{ik}x^i_\eta x^k_\eta
+ \frac{\partial }{\partial \alpha}
 \left(\frac{\rho}{4 \pi q}x^l_\alpha\right)+
 \frac{\rho}{4\pi q}\Gamma^l_{ik}x^i_\alpha x^k_\alpha=
 -\frac{g^{il}}{\rho q} \frac{\partial P}{\partial {x^i}},\hspace*{3mm}
\label{10}
\end{equation}
where $\Gamma^l_{ik}$ are Christoffel symbols.

The string equations (\ref{10}) for a flux tube embedded in a
gravitational field $g_{ik}(x^i)$ and a pressure field $P(x^i)$
can as well be derived from the action \cite{Semenov2000-123}
\begin{eqnarray}
S= -\int \frac{Q}{\rho}\sqrt{g_{ik}x^i_{\eta}x^k_{\eta}} d\eta
d\alpha. \label{13}
\end{eqnarray}
The action (\ref{13}) is invariant under $\eta$-reparametrisation
$\eta \rightarrow \eta'(\eta)$. Hence, the canonical Hamiltonian
vanishes identically and we need a gauge condition to fix the
parametrisation. The appropriate gauge condition is
\cite{Semenov2004-978}
\begin{equation}
q=\sqrt{g_{ik}x^i_{\eta}x^k_{\eta}}= \frac{1}{w}, \label{h5}
\end{equation}
where $w=\varepsilon+p/\rho$ is the enthalpy of the plasma. Using
the gauge condition (\ref{h5}) it can be shown that equations
(\ref{10}) are of hyperbolic type with relativistic Alfv\'enic and
slow mode  characteristics.

The Kerr metric in Boyer-Lindquist coordinates has
two cyclic variables, namely the coordinate time $t$ and the azimuth angle
$\phi$.  Therefore, there exist two conservation laws, for the string -
energy $E$ and the angular momentum $L$,
\begin{eqnarray}
E &=& \int_{\alpha_1}^{\alpha_2} \frac{Q}{w\rho}
(g_{tt}t_\tau+g_{t\varphi}\varphi_\tau) d \alpha, \label{consE}
\\
L &=& -\int_{\alpha_1}^{\alpha_2} \frac{Q}{w\rho}
(g_{t\varphi}t_\tau+g_{\varphi \varphi}\varphi_\tau) d \alpha.
 \label{consL}
\end{eqnarray}
It is supposed that there is no flux of energy and angular
momentum through the ends ${\alpha_1},{\alpha_2}$ of the flux
tube.

The string equations (\ref{10}) have been solved numerically using
the total variation diminishing (TVD) scheme. The conservation
laws (\ref{consE})-(\ref{consL}) have been used to control the accuracy
of the numerical scheme. More details on the method can be
found in \cite{Semenov2004-978}.
}

Summing up, a relativistic flux tube is characterised by the
internal parameters of density, pressure and magnetic field and it
is embedded in the a priori specified external gravitational field
and pressure field. Once this equation for the evolution of the
relativistic flux tube has been found, several properties from
classical MHD immediately generalise to general relativity: the
stretching of field lines leads to a decrease in density and,
consequently, to buoyancy forces; the infinite stretching of the
field line has to be stopped by some nonideal process like field
line reconnection~\cite{PriestForbes}.

\section{Energy extraction from a rotating black hole}
\begin{figure}
\centering
\includegraphics[width=\hsize]{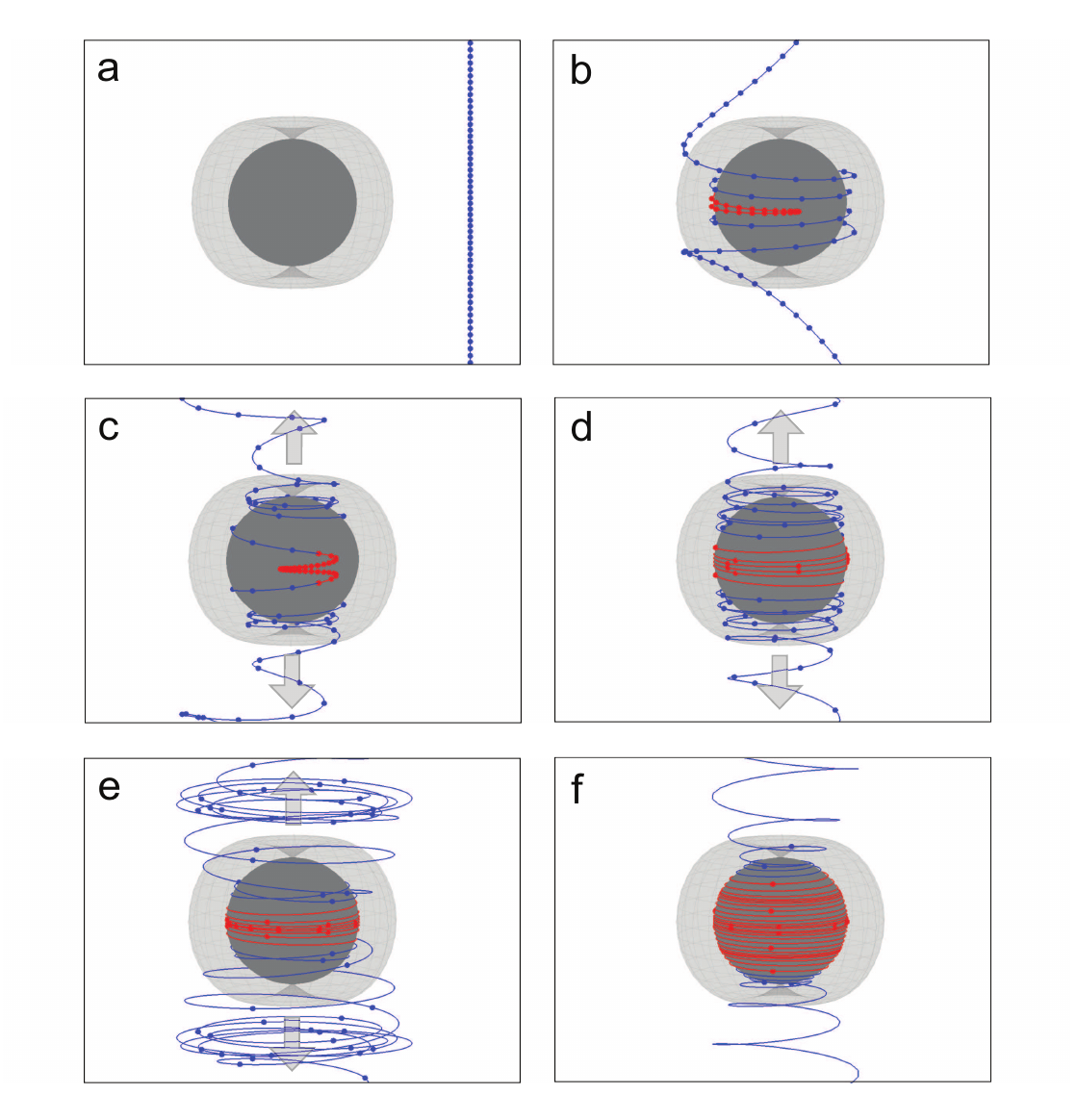}
\caption{Differential rotation due to the
Lense-Thirring effect near the Kerr black hole winds up the
massive magnetic flux tube (a-d). Inside the ergosphere (light grey),
magnetic tension slows down the rotation of the leading part of the
tube marked in red. This part carries negative energy and angular momentum.
Stretching of the flux tube is visualised by markers.
The stretched field line has lower density and buoyancy produces
the helical magnetic structure along the spin axis (e). Centrifugal
forces accelerate the plasma along the field lines.
This is the birth of the jet (f).
See Supporting Online Material: Animation 1.
}
\label{fig_1}
\end{figure}

\begin{figure}
\centering
\includegraphics[width=0.6\hsize]{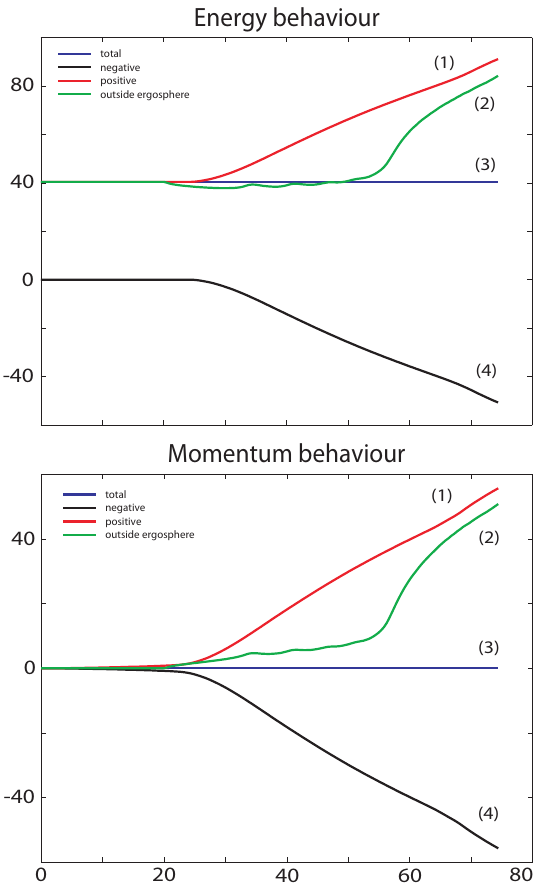}
\caption{Time history of energy and angular momentum.
The total energy (blue) of the flux tube is conserved.
Part of the tube inside the ergosphere loses energy, i.e.\ gains negative energy (black),
while other parts of the tube compensate this loss with positive energy (red).
Although the flux tube continuously falls into the ergosphere,
after some time $\approx50\,r_h/c$ with $r_h$ the radius of the event horizon,
the energy content of the flux tube outside the ergosphere (green) exceeds
the total initial energy of the flux tube (blue).
In the lower panel the corresponding angular momentum evolution is shown.
}
\label{fig_2}
\end{figure}

Consider an initially straight infinitely long
massive magnetic flux tube with zero angular momentum
everywhere along the string as sketched in Fig.~\ref{fig_1}a. If this
would be just a convected fluid tube (no magnetic field), each
part of the fluid tube would start to spin up with the Zero Angular
Momentum Observer (ZAMO) angular velocity as it falls into the
black hole. Due to the inhomogeneous character of the
Lense-Thirring torques, the flux tube becomes stretched and
twisted (Fig.~\ref{fig_1}b). The increase of magnetic tension slows down
the rotation of the central part of the tube nearest to the black hole
and therefore the latter will rotate locally slower than ZAMO
and thus gain negative energy and momentum. After a while negative
angular momentum (after $\eta\simeq 22$ in Fig.~\ref{fig_2}, lower panel) as
well as negative energy (a bit later, after $\eta\simeq 26$ in
Fig.~\ref{fig_2}, upper panel) appears in this leading part of the tube
(depicted in red in Fig.~\ref{fig_1}).
Since the total energy of the tube is conserved, some outer part
of the tube will have now more energy than it was initially
(Fig.~\ref{fig_2}, upper panel).
The whole process is similar to the Penrose
process but in the string mechanism there is no need for  the
decay of particles or tubes because angular momentum and energy
can be redistributed along the tube~\cite{Semenov2004-978}.
Animation~1 of the supporting online material shows the evolution of a field line falling into
the black hole and the corresponding distribution of energy and angular momentum.

In the string mechanism MHD waves are responsible for the
redistribution of angular momentum and energy. At the beginning of
this process the plasma flow inside the ergosphere is everywhere
radially directed inward, i.e. the plasma falls into the hole. It
is the MHD waves which can cross the static limit surface during
this initial stage and provide Poynting flux which can be observed
outside the ergosphere. In the course of time the process of field
line stretching continues, i.e.\ the magnetic field strength
increases whereas the density decreases. As a result, the speed of
the MHD wave strongly increases
and the angular momentum and energy transfer becomes even more
effective.

\section{Buoyancy}

At this stage, another physical process plays an important role:
the part of the flux tube with extra positive energy looses more
and more plasma and the flux tube will feel the relativistic
analogon of the buoyancy force~\cite{Parker1979}. 
\Red{Unfortunately it is not possible to find such a distribution of
total pressure which can balance the gravity of a Kerr black hole, 
because even
without a magnetic field there should be a plasma flow across 
the event horizon. 
There exists no a simple solution to this problem and we can
not use the distribution of the plasma pressure in the right hand
side of the string equation (\ref{10}). But the Kerr black hole 
is such a powerful object that the actual distribution of the total
pressure is not that important. In fact, the buoyancy force}
first slows
down the radial plasma accretion and then eventually pushes some
fragment with positive extra energy along the spin axis outside
the static limit surface as shown in Fig.~\ref{fig_1}c-~\ref{fig_1}e. This is
the birth of a jet (see Fig.~\ref{fig_2} at $\eta \simeq 50$). The
buoyancy force generates the pronounced double helical
magnetic field structure aligned with the spin axis.
Along these field lines, the plasma is centrifugally accelerated
to nearly the speed of light.

An important property of this mechanism including buoyancy is that
it does not depend on the detailed initial magnetic field distribution
except that the characteristic size of the magnetic field should be
of the order of the event horizon. In this case, the Lense-Thirring
effect produces the spiral magnetic field while buoyancy creates the
outgoing helical structure.
Animation~2 of the online supporting material shows the same as Animation~1
but for an inclined magnetic field line.

This process evidently leads to an unlimited stretching of the
flux tube since one part of the string continues to fall into the
hole and simultaneously another part of the string is pushed
outward. \Red{An important fact is that this is a continuous 
process which does not depend sensitively on actual distribution 
of the total pressure.} Apparently at some time this stretching must be limited
by some nonideal process. Two such mechanisms are discussed in the
literature. The first mechanism is popular in astrophysics and
considers a gap with an extremely strong parallel electric field
such that this field can produce electron-positrons pairs from the
electromagnetic field. Such a gap will produce an
electron-positron plasma which will fill up the tube. One can
imagine such a process even steady state and axisymmetric. The
physics of the steady state process is based on the
Blandford-Znajek mechanism further discussed in
~\cite{Punsly1990-518,Punsly2001,Beskin2009,Narayan2005-199}.

\begin{figure}
\centering
\includegraphics[width=0.8\hsize]{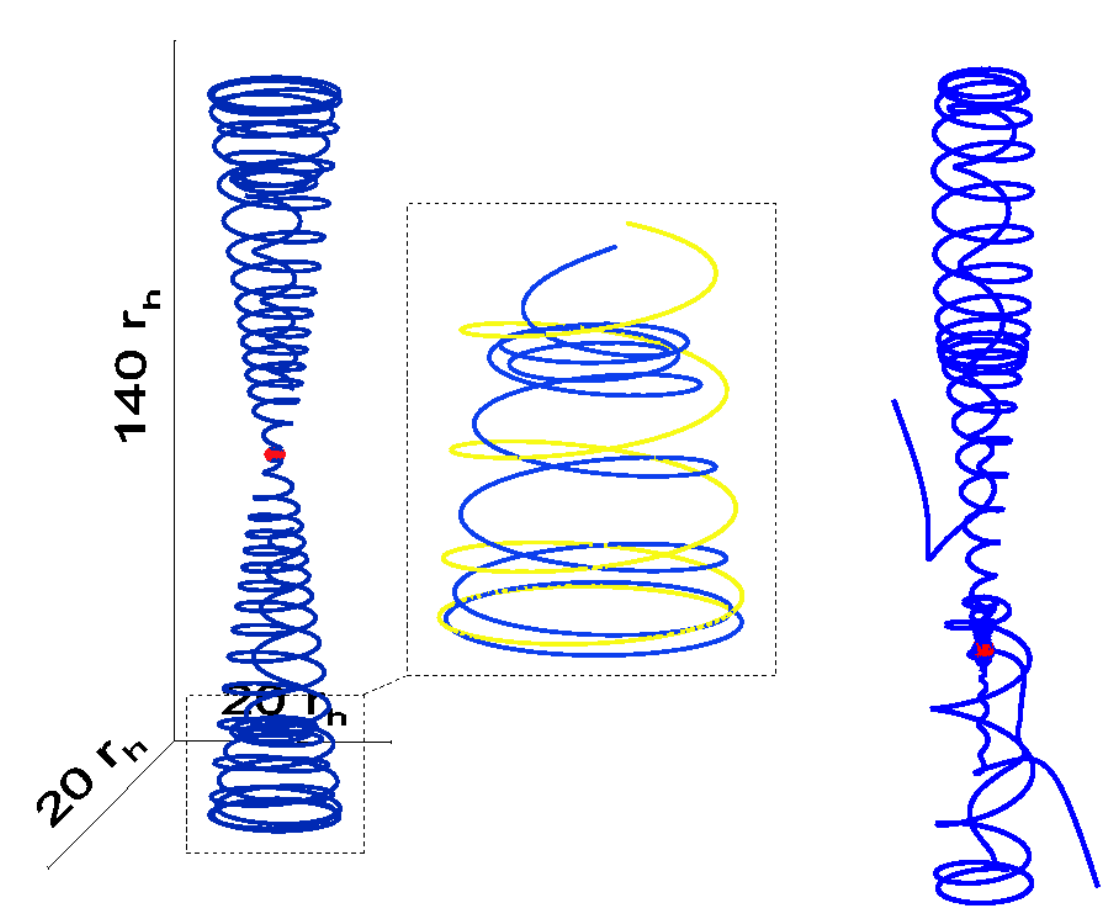}
\caption{Differential rotation twists the magnetic field which is then
expelled by the buoyancy force along the spin axes and forms a
double helical structure.
On the left the evolution of a field line initially parallel to the spin
axis, on the right the evolution of a field line initially inclined
with respect to the spin axis. Although the structure is asymmetric for the
inclined case, the resulting helical structure is still along the spin axis.
Magnetic reconnection shown in the enlarged detail (left) creates
autonomous magnetic structures (plasmoids) detached completely from the
black hole.
See Supporting Online Material: Animation 2.
}
\label{fig_3}
\end{figure}

A second possibility to limit the stretching process is magnetic
reconnection. It consists of a local breakdown of the ideality
condition which allows for a slipping of magnetic field lines with
respect to the plasma flow. In this nonideal regions the field lines
are topologically reordered~\cite{Blackman1996-L87,Semenov2004-978,Giannios2009-L29}.
Most effectively this will work
if the stretched flux tube is reconnected outside the static limit
boundary to itself (Fig.~\ref{fig_3}).
A closed double helical structured field line with low density will
form a bubble which will freely evolve and the release of magnetic
tension will power the jet stream while the
ergospheric part of the tube is supplied with new accreting material.
Animations~3 and 4 of the supporting online material show details of this evolution.

\section{Jet Relaxation}
\begin{figure}
\centering
\includegraphics[width=0.8\hsize]{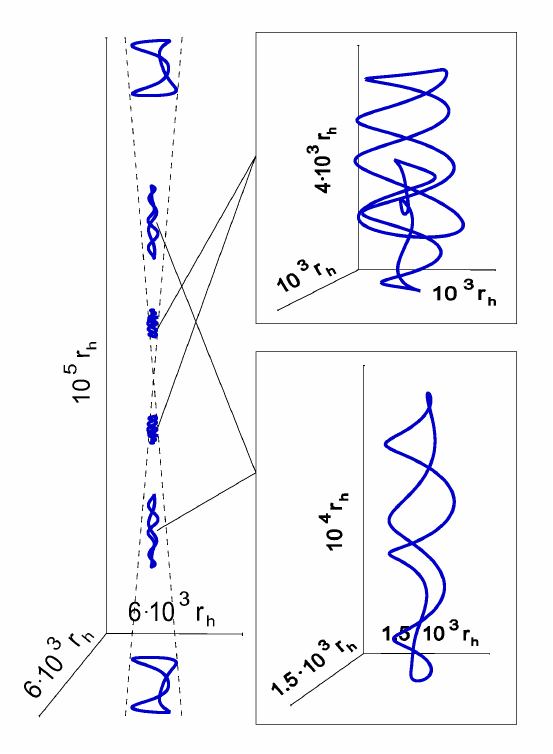}
\caption{Reconnection outside the ergosphere produces magnetic bubbles with helical magnetic
fields and a rotating plasma inside. Each plasmoid carries away rotational energy
of the black hole similar to the outgoing particle in the Penrose mechanism.
The evolution of such a plasmoid is shown for different time steps.
The helical magnetic field relaxes into a simpler circular structure and,
as a result, the collimation of the jet is lost in the final stage of evolution.
See Supporting Online Material: Animation 3 and 4.}
\label{fig_4}
\end{figure}

The structure of the magnetic field inside the outgoing plasma bubble
becomes more and more simple due to the relaxation of Maxwellian stresses
and the double helical structure will relax into a simpler circular structure.
As a consequence, the collimation effect becomes less and the width of the jet
slowly increases. At some time, the magnetic field cannot confine the
rotating plasma any more, and the jet quickly spreads out (Fig.~\ref{fig_4}).
In this picture, there is no need for collisions with external plasmas.
Each plasmoid can be considered like an outgoing particle in the
Penrose mechanism: it carries additional positive energy away from the
black hole.
Therefore, the Penrose mechanism is used in a two-fold sense in this scenario,
first, for the behavior of the flux tube inside the ergosphere, and, second,
for the behavior of the plasmoid outside the ergosphere.

\Red{So far we considered only self-reconnection of magnetic
flux tubes which produces relatively simple structures of the magnetic
field. In reality the situation is expected to be more complicated and
chaotic: some tubes fall down into the black hole and simultaneously the
other tubes will rise due to buoyancy. Hence, they can collide and
reconnect with each other. In other words, the probability for a flux tube 
to reconnect with some other tube seems to be higher than
self-reconnection. In such a case, the magnetic structures will be
very complicated and during the relaxation stage we can expect
additional collisions of the outgoing flux tubes. As a result,
further reconnection events are likely to occur also during the next stage 
of jet propagation. Reconnection converts magnetic energy into 
kinetic and internal energy of the plasma and give rise to
energetic particles which is important for models of
extragalactic radio sources ~\cite{Blackman1996-L87}.

\section{Conclusions}
Buoyancy effect can play an important role in the course of jet
creation. At the beginning it slows down the motion of the plasma
towards the black hole, especially for that part of the tube with
low density. Eventually this part starts to move outwards
producing a spiral structure needed for centrifugally driven
jets. This might resolve an important questions of relativistic jets:
how does a magnetised plasma falling into the huge gravitational
center of a black hole which, in principle, swallows everything,
produce outgoing streams near the speed of light.

A further necessary element in this scenario is magnetic field line
reconnection.
In this context the question arises where is the correct place
to reconnect the field lines. One might expect that this place
is near the event horizon because of nearly antiparallel field
lines there.
However, it turns out that
gravity in this region is so strong that the release of Maxwellian
stresses there by reconnection cannot prevent the plasma from 
falling into the black hole. The more appropriate site
for reconnection is likely to be near the static limit surface, where
magnetic reconnection can produce a chain of plasmoids and supply 
the black hole with new material.

In addition, we propose a new relaxation mechanism. 
It is well known that jets eventually start to spread out,
getting wider and wider. Usually this is connected to the collision 
with the plasma outside or a Kelvin Helmholtz instability. 
Our point of view is that Maxwellian relaxation continues
all the way long and leads to a simplification of the magnetic jet 
structure. Roughly speaking, each plasmoid tends to become a circle-like 
structure which leads to a jet of finite  thickness.

In our scenario the  Penrose mechanism has been used twice.
Locally to redistribute angular momentum and energy
along the string and, consequently, to extract energy from the
Kerr black hole, and globally in form of an outgoing plasmoid which
transfers additional positive energy and angular momentum away
from the black hole. This plasmoid can be considered as a
classical Penrose particle which carries away rotational energy 
from the black hole.

The accretion of magnetised plasma into a Kerr black hole with
differential rotation results in a common pattern, i.e.\ a helical
structure of the magnetic field, a spinning plasma flow which
eventually leads to jet formation via the frame dragging effect,
the Penrose mechanism, relativistic buoyancy, centrifugal
acceleration and reconnection. The predictions of the present
mechanism results from the application of energy and angular
momentum conservation for magnetic flux tubes. There is
practically no dependence on the initial configuration of
the magnetic field.
}

\ack
The authors thank B.~Punsly for useful discussions.
       VSS acknowledges financial support from the TU Graz.

\section*{References}
\bibliography{Rel_Jet}
\bibliographystyle{aip}

\clearpage

\clearpage

\end{document}